# Serverless Computing for Scientific Applications

Internet Computing - View from the Cloud Section

Maciej Malawski[1,2], Bartosz Balis[2]

1. Sano Centre for Computational Medicine, Krakow, Poland (https://sano.science)
2. AGH University of Science and Technology, Institute of Computer Science, Krakow, Poland

Abstract

Serverless computing has become an important model in cloud computing and influenced the design of many applications. Here, we provide our perspective on how the recent landscape of serverless computing for scientific applications looks like. We discuss the advantages and problems with serverless computing for scientific applications, and based on the analysis of existing solutions and approaches, we propose a science-oriented architecture for a serverless computing framework that is based on the existing designs. Finally, we provide an outlook of current trends and future directions.

## Introduction

Given the increasing role of simulations and data analysis in science today, researchers never have too much computing power. For this reason various dedicated research computing infrastructures are built, including HPC centers, large-scale computing clusters or grids, or smaller centers for research computing at universities and research institutes. On the other hand, distributed computing in the industry leverages large-scale datacenters, which provide computing resources based on the cloud computing model. Over the last 10 years, these commercial offerings in the form of public clouds have been of interest to the scientific community, and the development of cloud solutions influenced the way traditional HPC hardware and software have evolved. These technological trends initially have included virtualization, containerization, on-demand access to resources, or object storage services. With increasing cloud adoption, advanced *cloud-native* technologies have emerged, with the Kubernetes container orchestration technology being their cornerstone [1].

Recently, we observe a new trend in cloud technologies, which is generally called "serverless computing". In general, serverless computing allows executing functions with minimum overhead in server management, combining developments in microservice-based architectures, containers and the new cloud service models such as Function-as-a-Service (FaaS) and Container-as-a-Service (CaaS) [2], [3] .

The serverless computing model has not been designed to support scientific computing, rather it has targeted lightweight event-based applications. Still, as the research community is very open to exploring new and alternative ways of accessing computing resources and building scientific applications, we can see many attempts to evaluate the applicability of the serverless model for scientific applications and the desire to repurpose it to the requirements of the scientific community.

In this paper, we provide examples of using serverless model for scientific applications, based on our experience in this area. This provides our perspective on how the recent landscape of serverless computing for scientific applications looks like. We discuss the advantages and problems with serverless computing for scientific applications, and based on the analysis of existing solutions and approaches, we propose a science-oriented architecture for a serverless computing framework that is based on the existing designs. Finally, we provide an outlook of current trends and future directions.

# Strengths and weaknesses of serverless model for scientific applications

There are several benefits of using serverless model or FaaS:

- **Function as a useful abstraction**: functions are the most fundamental abstractions in mathematics, which is called the language of science. Functions are also a very powerful abstraction in computer science, with theoretical foundations in the lambda calculus and the functional programming paradigm. Functional programming has proven to be useful for distributed systems, with the examples of Erlang and Scala languages and actor systems implemented in them [4]. In scientific computing, modern programming languages such as Julia are also strongly influenced by the functional programming paradigm. For these reasons, FaaS offers a natural abstraction for scientific applications using distributed computing.
- **Simplicity facilitating programming**: the general premise of serverless computing is to facilitate application programming by hiding the complexity of underlying infrastructure and relieving the programmer or user from managing the infrastructure. All the intertwined resource management issues such as provisioning, scheduling or autoscaling are in principle handled by the provider in a much more broader scope than in any other cloud service model or in any distributed programming environment.
- **Highly-elastic resource management model**: in serverless computing, the unit of resource allocation is a single function call, and the FaaS platforms are designed for serving large amounts of fine-grained requests very quickly. Our experiments have shown that it is possible to request thousands of concurrent function calls and they are invoked in parallel within seconds by a cloud provider such as AWS or Google [5]. The overhead is really low as compared to IaaS clouds (minutes) or in HPC systems (hours). This opens the possibility of better support for interactive and dynamic workloads, which are of interest for scientific applications.
- **Deployment model using familiar programming languages and containers**: while FaaS was originally designed for Web or mobile applications based on JavaScript, Java or Python, now it is possible to add custom language support or container images such as Docker. This is fundamental for scientific computing, where applications typically require diverse programming languages, libraries and tools.

Another highly popular approach for scientific computing in the cloud is using a *container orchestration platform*, with Kubernetes being a de facto standard, supported by all major cloud providers. Kubernetes can be seen as a middle ground between plain Infrastructure-as-a-Service cloud and FaaS. On the one hand, Kubernetes hides the complexity of cluster management, on the other hand application development in Kubernetes still requires significant IT engineering skills. In our research, we have investigated various aspects of scientific workflow management in Kubernetes. This perspective lets us point out a few problems of FaaS in the context of scientific computing and contrast them with Kubernetes.

- **Vendor lock-in**: serverless code usually uses various cloud services through vendor APIs which increases the chance of vendor lock-in. Developing a portable solution in FaaS is harder than in Kubernetes.
- **Observability**: with platform and infrastructure hidden behind APIs, system observability is out of control of the developer. However, observability is the cornerstone of experimental science. Diagnosing problems and getting important metrics related to performance, energy consumption, etc., can be more difficult in FaaS than in Kubernetes where the developer has full control over the observability stack.
- **Economics and performance**: the serverless model is very attractive for production systems that run 24/7 and need to handle variable workload. In such cases, a possibly higher per-cycle cost of FaaS can be mitigated by high elasticity. However, for a scientist who runs one-off batch workloads this is not necessarily the case. It has been shown that for data-intensive workloads, such as model training, FaaS is considerably more expensive and slower than IaaS [6]. *Performance isolation* is another challenge. E.g. in [7], the best performance was achieved on IaaS by allocating one CPU to one computational task. Such control over resource management, also possible in Kubernetes, is not available on FaaS, where the resources are managed by the underlying platform [8].

As we can see, there are numerous potential benefits and challenges of using serverless computing for scientific applications. In the next section we show examples of how these have been addressed in the specific solutions targeting scientific computing.

# Examples of scientific applications and frameworks using FaaS

From the beginning of the serverless model and from the first releases of FaaS services, they have been noticed as potential sources of computing for scientific applications. Here we provide a set of selected examples, which we think nicely represent typical scenarios.

## PyWren

Perhaps the first framework for running compute-intensive workloads on serverless platforms which received wider popularity was PyWren [9]. As a simple, yet powerful Python library, it allows running stateful functions in parallel, using shared cloud storage for input and output, similarly to a tuple space model proposed in Linda. An interesting technical solution in PyWren is to use Cloudpickle Python library, which allows one to serialize and execute remotely arbitrary Python code. Cloudpickle was developed by PiCloud.com start-up company, which offered simplified computing based on Python functions invoked on AWS cloud more than 5 years before FaaS model was proposed. PyWren has been applied to many embarrassingly parallel problems such as MonteCarlo simulations, parameter sweeps, etc, but also for MapReduce style data processing tasks, video encoding, parameter optimization for distributed machine learning or distributed compilation.

## FaaSification of scientific applications

One of the first discussions of the potential for using FaaS for scientific computing was presented by Spillner et al [10]. The authors describe four experiments which compare the performance and resource consumption for example benchmarks or applications: calculation of π, face detection, password cracking, and precipitation forecast. The experiments are run using AWS Lambda and a local testbed using Snafu tool developed by the authors. In addition to performance evaluation, there is a very interesting discussion about possible strategies for FaaSification (adaptation of existing software to FaaS) of monolithic applications, which can be done at varying levels of granularity: from whole functions to single lines of code. This discussion brings an important topic of the effort needed to make use of the serverless computing model to existing applications and the potential for using tools for automation.

## Serverless scientific workflows

FaaS can be a good fit for scientific workflows (graphs of tasks), in particular those with large numbers of relatively fine-grained tasks. HyperFlow, our workflow engine developed at AGH, was extended to FaaS platforms [11], including Google Cloud Functions, AWS Lambda, and other functions based on HTTP request interface. In HyperFlow, stateless functions operate in a download-compute-upload sequence, using cloud storage for input and output. Initially we had to work around deployment problems by building custom-compiled binaries compatible with the operating systems of the FaaS providers, but recent support for Docker images solved this problem. Similarly, Container-as-a-Service serverless platforms, such AWS Fargate and Google Cloud Run, proved to be a viable solution for scientific workflows [12]. Other interesting examples include Triggerflow [13], an event-driven workflow framework based on triggers, and Abstract Function Choreography Language (AFCL), which offers high-level notation for workflows with a rich set of control- and data-flow constructs [14]. In our opinion, support for serverless backends will become a natural evolution of scientific workflow engines.

## NumPyWren

Seemingly, it is hard to imagine using serverless platforms for dense linear algebra, a domain traditionally reserved for HPC. Nevertheless, as NumPyWren and LAmbdaPACK [15] tools show, algorithms such as matrix multiplication or decomposition, can run efficiently in the cloud. NumPyWren uses cloud object storage for communication, and while the latency of Amazon S3 is orders of magnitude higher compared to MPI-over-Infiniband, the aggregate bandwidth and its scalability allows efficiently decomposing the matrix calculations into basic operations on tiles of such size that the high latency is compensated by high bandwidth. While the experiments show that the performance achieved does not immediately beat the MPI implementation, the benefits of serverless approach are scalability, elasticity and better resource utilization. Notably, the framework wisely combines various additional cloud services: SQS for task queue and Redis or DynamoDB as key/value store for managing the state.

## ROOT Lambda - serverless tools for High Energy Physics

The High Energy Physics community, having a long tradition of leveraging distributed computing infrastructures, shows a growing interest in exploring modern frameworks coming from the big data industry. A notable example includes Distributed RDataFrame, an extension to ROOT framework for data analysis adding the high-level functionality based on the Data Frame model. RDataFrame compiles operations such as filters and aggregations into a graph of tasks, and supports multiple backends, including local multiprocessing, Apache Spark, Dask, and recently AWS Lambda [16]. When developing the AWS backend, we solved many technical problems, including containerized deployment of ROOT and remote access to storage at CERN. Despite the relatively large volume of data transfer between CERN and AWS, this approach scales to at least hundreds of parallel Lambda tasks, allowing interactive data analysis. This exemplifies the potential of using serverless infrastructures as backend to domain-specific scientific tools by providing user friendly abstractions.

## FuncX

A custom-developed solution which aims specifically at supporting scientific applications using the serverless model is FuncX [17]. It supports running FaaS applications on federated resources ranging from local computers, via clusters and clouds to supercomputers, with focus on applications with fine grained tasks. Examples of such applications are scalable metadata extraction, machine learning inference, crystallography, neuroscience, correlation spectroscopy, and high energy physics. The interesting feature of the approach is that it does not simply provide access to FaaS platforms, but brings together resources from multiple sources, including those dedicated to scientific computing and e.g. equipped with specialized hardware such as GPU. The example applications of FuncX show that scientific computing often relies on many tasks which do not necessarily require a traditional supercomputer.

## High-throughput biomedical application examples

High-throughput computing is often required in biomedical applications for screening of large space of molecular configurations. Examples in proteomics are Replica Exchange Molecular Dynamic (REMD) which have been successfully ported to serverless computing [18]. The usage of serverless architecture allows for more dynamic scaling of the workers (executed on FaaS), while it requires adding a communication layer using cloud object storage or Redis database. A similar approach was used in serverless implementation of Smith-Waterman dynamic programming algorithm for comparing protein sequences [19]. A recent survey [20] shows several examples of various applications of serverless computing to omics data analysis and integration, all representing high-throughput architecture with a scalable pool of resources obtained using the FaaS or CaaS model.

## Distributed Machine Learning

When discussing scientific or computational applications, one cannot exclude training and serving of Machine Learning models, which can also be a subject of porting to serverless architectures. One of the early examples [21] uses AWS Lambda for inference of large neural network models. SIREN [22] is a distributed framework running compute-intensive batches of ML tasks on FaaS. Another example is FedLess [23], which is a serverless framework for secure training of ML models using a federated learning approach.

# Layered ecosystem of applications and generic architecture of the frameworks

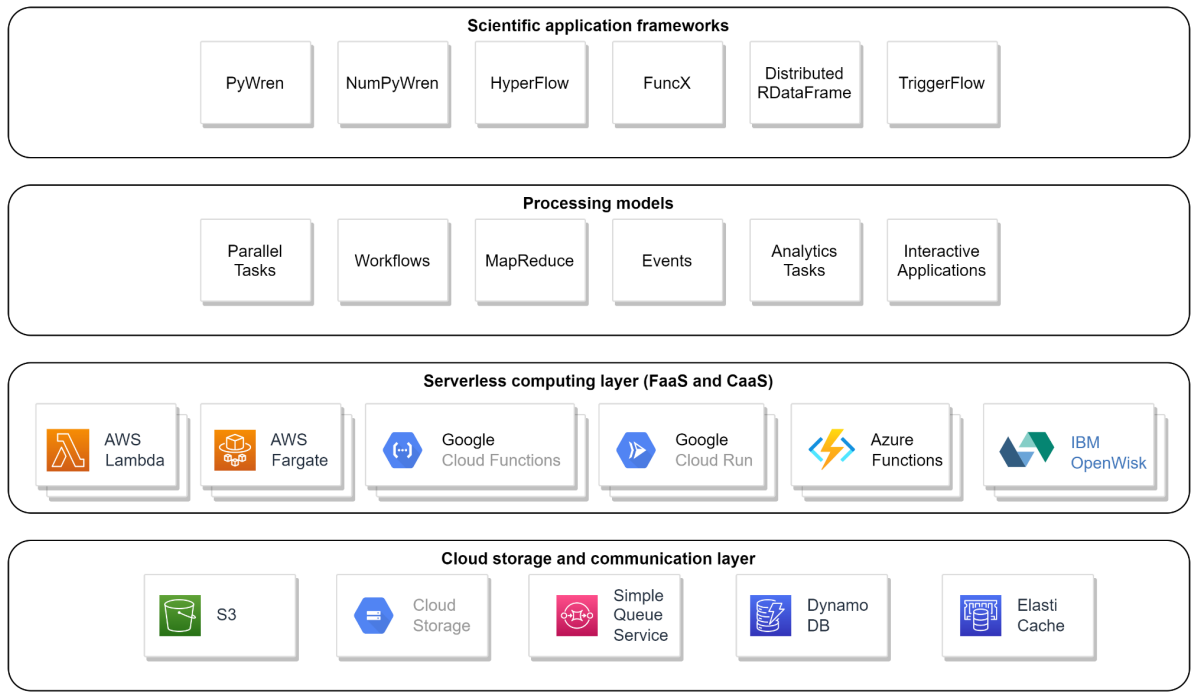

Figure 1 Layered ecosystem of serverless scientific applications.

Having studied examples of serverless scientific applications, we can observe an emerging layered architecture of their ecosystem, as shown in Figure 1. From the bottom up, we have the basic layer of cloud storage and communication, which includes cloud object storage, queue systems or caches. This layer provides state management for the stateless FaaS/CaaS layer. Next come various processing models - each of them relevant for scientific users and software engineers. Finally, the top layer includes ready to use frameworks for scientific applications, which typically provide high-level APIs or user interfaces.

Another general observation is that most of the frameworks have a very similar common architecture, shown in Figure. 2. The main component, called Execution Engine, is responsible for task orchestration and it uses some form of database, typically a Key-Value Store, for managing the internal state of the application. The tasks are submitted to a Task Queue and then are processed using stateless FaaS or CaaS services, which use Cloud Storage for data exchange. There are possible variations of this architecture, including more distributed or decentralized orchestration, some frameworks do not use any queue, but directly invoke FaaS or CaaS functions using the public API, finally there are multiple options regarding database and cloud storage backends (see Figure 1). Nevertheless, this typical architecture can be considered as a standard blueprint for building computing frameworks using the serverless model.

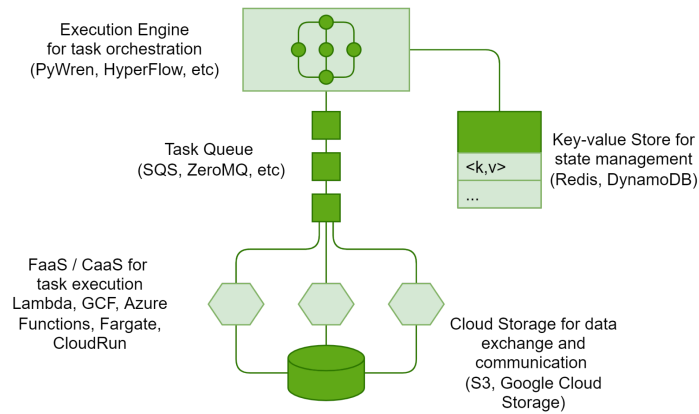

Figure 2 Generic architecture of a serverless execution framework for scientific applications.

## Summary and outlook

The examples presented here show that the concept of serverless computing can be applied in scientific computing, where traditional distributed processing or high-throughput approaches have been used. The new capabilities offered by serverless, including simplified programming model based on functions, highly-elastic resource management and convenient deployment model, allow not only for repurposing of existing applications and frameworks (parallel tasks, workflows, MapReduce, etc), but can inspire new classes of scientific applications, which can be more event-driven, interactive and highly dynamic in resource usage, and also take advantage of the whole continuum of resources from HPC, cloud and other devices located at the edge.

There are of course limitations of serverless computing, such as vendor lock-in, observability issues, cost-performance trade-offs, distributed state management, caching, and lack of tooling - these topics are now subject of active research [3]. Some trends such as datacenter disaggregation [15], convergence between HPC and cloud architectures, and increasingly elastic resource management in clouds may suggest that some form of "serverless" computing will become prevalent. The future will show if a "serverless supercomputer" may become an ultimate solution to scientific computing problems, but at least we are certain that the concepts presented here will influence the future developments in both compute infrastructures and the architectures of scientific applications.

## Acknowledgments

This work was supported by the EU H2020 grant "Sano" No 857533; and by the project "Sano" carried out within the International Research Agendas Programme of the Foundation for Polish Science, co-financed by the European Regional Development Fund.